\documentclass[12pt]{article}
\usepackage{a4}
\usepackage{epsf}
\usepackage{amssymb}
\usepackage{cite}
\newcommand{\be}{\begin{equation}}
\newcommand{\ee}{\end{equation}}
\newcommand{\bea}{\begin{eqnarray}}
\newcommand{\eea}{\end{eqnarray}}

\begin{document}
\def\C{{\mathbb{C}}}
\def\R{{\mathbb{R}}}
\def\s{{\mathbb{S}}}
\def\T{{\mathbb{T}}}
\def\Z{{\mathbb{Z}}}
\def\W{{\mathbb{W}}}
\def\Bbb{\mathbb}
\def\BZ{\Bbb Z} \def\BR{\Bbb R}
\def\BW{\Bbb W} 
\def\BM{\Bbb M} 
\def\e{\mbox{e}}
\def\BC{\Bbb C} \def\BP{\Bbb P}
\def\CP{\BC\BP}
\begin{titlepage}
\title{On Tachyons in Generic Orbifolds of $\BC^r$ and Gauged Linear Sigma
Models}
\author{}
\date{
Tapobrata Sarkar
\thanks{\noindent E--mail:~ tapo@iitk.ac.in}
\vskip0.4cm
{\sl Department of Physics, \\
Indian Institute of Technology,\\
Kanpur 208016, India}}
\maketitle
\abstract{We study some aspects of Gauged Linear
Sigma Models corresponding to orbifold singularities of the form 
$\BC^r/\Gamma$, for $r=2,3$ and $\Gamma = \BZ_n$ and 
$\BZ_n\times \BZ_m$. These orbifolds might be tachyonic in general. 
We compute expressions for the multi parameter sigma model 
Lagrangians for these orbifolds, 
in terms of their toric geometry data. Using this, we analyze
some aspects of the phases of generic orbifolds of $\BC^r$. 
}
\end{titlepage}

\section{Introduction}\label{intro}

Tachyon dynamics have been studied extensively in the last few
years, following the pioneering work of Sen \cite{sen}, and is
important for the general understanding of the role of time in 
string theory. Whereas open string tachyon condensation processes
lead to the decay of D-branes, the analogous condensation of closed 
string tachyons (that breaks space-time supersymmetry) 
result in a decay of spacetime itself. The general 
problem of studying the dynamics of tachyons in closed string
theories is quite difficult, since there might be delocalised tachyons
in the theory, but a class of theories where the problem is more
tractable are theories with localised closed string tachyons. 
This was the issue addressed by Adams, Polchinski and Silverstein
(APS) \cite{aps}. 

The approach of APS was to study the brane probe picture of string
theories with localised tachyons, which arise as twisted sector states
in the corresponding closed string theory. Clearly, the problem involves
two distinct scales. In the substringy regime, where 
$\alpha'$ corrections are small, D-branes probing the orbifold
in question provide a good description to the singularity structure
via the gauge theory living on the brane. Far from the substringy regime, 
the brane probe picture is less useful and one has to resort to
a supergravity analysis. In any case, in the open string picture,
one finds that by exciting marginal or tachyonic deformations in the theory,
one can drive the original orbifold to one of lower rank, and
tachyonic deformations of the latter takes the system to a 
final stable (supersymmetric) configuration \footnote{This is 
strictly true only for $\BC/\BZ_n$ and $\BC^2/\BZ_n$ orbifolds.
For $\BC^3/\BZ_n$ orbifolds, in the absence of a canonical 
resolution, there might be terminal singularities, i.e the
end point of tachyon condensation need not result in a 
singularity that can be resolved solely by marginal deformations.}  

An useful alternative way to study the decay of localised closed string
tachyons is to focus on the $N=(2,2)$ CFT of the worldsheet.
By using an appropriate Gauged Linear Sigma Model (referred to
as the GLSM in the sequel), one can effectively track down the 
decays of closed string tachyons \cite{vafa}\cite{hkmm}
\cite{mm} (for reviews, see \cite{martinec},\cite{mt}).  
For orbifolds of the form $\BC^3/\BZ_n$, this analysis was
carried out in \cite{mnp},\cite{ts1}, and it was shown in 
\cite{mnp} that generically, while the decay products of Type II
string theories with tachyons in the spectrum can be 
deformed to flat geometries using only marginal deformations
of the resulting theory at the endpoint of the tachyon decay, Type 0 
theories might have terminal singularities as the endpoint of
the tachyon decay process. Further, in \cite{mn}, the GLSM was used to
effectively study the phases of the theory, (for two and three 
parameter GLSMs), and a rich phase structure of the GLSM was found. 
In \cite{ts1}, a slightly different approach to the problem of
tachyon condensation in the non-supersymmetric $\BC^2/\BZ_n$ and 
$\BC^3/\BZ_n$ was followed, and the classical sigma model metrics
for the GLSMs were calculated, which effectively described the 
condensation of localised closed string tachyons in the same. Here,
a single parameter GLSM (with a single $U(1)$ gauge group) 
was considered, corresponding to the most
relevant tachyon (i.e the one with the highest (negative) mass
squared). Condensation of the most relevant tachyon of course leaves
the possibility of then exciting other tachyonic modes (which remain
tachyonic at the end of the decay).   

In general, the GLSM describing the geometry of the instability
due to closed string tachyons will be charged under multiple 
$U(1)$ gauge groups, and it is important to have an full 
understanding of the same, in order to understand the phase 
structures of these, in lines with \cite{mn}. An advantage of doing
this is that one can then study the phase structure of arbitrary
charge GLSMs, without resorting to a case by case analysis. Further,
the non-linear sigma model metrics for these unstable geometries 
can possibly be understood via a generic toric description. It is
this issue that we set out to address in this paper. In particular,
we will show how to construct the bosonic Lagrangian for generic
multi parameter GLSMs with arbitrary charges, entirely in terms of the 
toric data of the target space singular geometry that the GLSM describes.
This will help us to read off the phase structures of the 
corresponding theories and their classical metrics. On the other hand,
D-brane dynamics in generic orbifolds of $\BC^r$ can be understood
in terms of open string GLSM boundary conditions, as shown 
in refs. \cite{hiqv}, \cite{tapo3}. As we will see later, 
our results here can be used to extend the analysis of these papers to
arbitrary charged GLSMs, and might be important in the study of
D-brane dynamics in such models.  
extend the analysis of.   

The paper is organised as follows. In section 2, we review the toric
construction of singularities of $\BC^2$ and $\BC^3$, in 
order to set the notation and conventions used in the rest of the paper.
In section 3, we first construct the bosonic Lagrangian for
one and two parameter GLSMs, and then move on to construct the 
Lagrangian for a generic multi-parameter GLSM with arbitrary charges.
In section 4, we apply the construction of section 3, to study some
aspects of the phases of generic orbifolds of $\BC^3$. Section 5 
contains our conclusions and discussions.  
 
\section{Closed String Tachyons on $\BC^r/\Gamma$}

In this section, we state some of the relevant results for closed
string tachyon condensation of orbifolds of the form $\BC^r/\Gamma$,
for $\Gamma = \BZ_n$ or $\BZ_m\times \BZ_n$. 
This section is primarily review material, in order to set the notations
and conventions used in the remainder of the paper. 

String theory on the space-time non-supersymmetric orbifold 
$\BR^{7,1}\times \BC/\BZ_n$ with the quotienting group acting on a 
complexified direction $Z$ with the action
\begin{equation}
Z\to \omega Z;~~~~~ \omega=e^{\frac{2\pi i}{n}}
\end{equation}
was studied by Adams, Polchinski and Silverstein (APS) 
\cite{aps}. The closed string conformal field theory is tachyonic in 
all its twisted sectors, and localisation of the tachyons demands that
$n$ be odd in Type II theories. The world volume gauge theory of a
D-p brane probing this singularity (the brane being located at the orbifold
fixed point) can be constructed a'la Douglas and Moore \cite{dm}, and 
the orbifolding action will retain only those fields invariant under
its action, thus resulting in a quiver gauge theory on the D-brane
world volume. APS showed that by giving vevs to certain scalar fields in the
resulting gauge theory, some of the other fields become massive 
(as can be determined from the scalar
potential of the gauge theory), and the remaining massless fields 
describe the quiver gauge theory of a lower rank orbifold. Fermionic
fields in the gauge theory are analysed similarly, using the Yukawa
coupling terms. The probe brane analysis is valid at substringy regimes,
and when gravity effects become large, one has to resort to 
supergravity techniques. The two yield a consistent description of 
the decay of the conical singularity into flat space-time, as
demonstrated by APS (see also \cite{headrick}, \cite{gh}). 

The case of complex two fold orbifolds yields a richer structure. In
this case, we consider string theory in the background 
$\BR^{5,1} \times \BC^2/\BZ_n$, where $\BR^{5,1}$ is flat six dimensional
Minkowski space-time, and the remaining four directions are 
complexified, with the orbifolding group acting as
\be
\left(Z^1, Z^2\right) \to \left(\omega Z^1, \omega^k Z^2\right),
~~~~\omega=e^{\frac{2\pi i}{n}}
\ee
this orbifold action (denoted by $\BC^2/\BZ_{n(k)}$) breaks space-time 
supersymmetry, whenever $1+k \neq 0~{\mbox{mod}}~n$. Again the open 
string analysis in the substringy regime can be carried out in the same 
way as in the $\BC/\BZ_n$ case. The APS procedure here is to give vevs
to some of the scalar fields by first turning on marginal deformations
in the theory that breaks a part of the original orbifolding group, 
and takes the theory to a locally supersymmetric, lower rank orbifold. 
Deformations of the latter, which are tachyonic, then drives the system
to a supersymmetric background. The supergravity analysis in this case
is more complicated than in the case of the $\BC/\BZ_n$ orbifold, and
one has to possibly resort to numerical techniques in order to study
the same. A similar analysis can be carried out for $\BC^3$ orbifolds
as well.  
  
In \cite{tapo2}, a brane probe analysis was carried out for the 
non-supersymmetric orbifolds of APS, following the procedure of 
\cite{asp}, \cite{dgm}. It was found that the procedure of
\cite{dgm} yields the correct toric geometry (see refs
\cite{fultonoda}, for comprehensive introductions to
toric varieties) from the gauge theory of D-branes 
probing non-supersymmetric orbifolds, but interestingly with 
certain marginal directions being turned on. This complemented
the work of \cite{vafa}, \cite{hkmm}, \cite{mm}, where an equivalent 
picture of the APS decay process was given in terms of the 
$(2,2)$ world sheet superconformal field theory (SCFT) of closed strings on
non-supersymmetric orbifolds, which is closely related to the toric
description of such orbifolds, and in turn relates directly to the gauged
linear sigma model of Witten \cite{wittenphases}. The open string
description in lines with \cite{dgm} becomes technically complicated for 
generic orbifolds, and in this paper we will primarily focus on the
equivalent closed string description.   

The closed string SCFT description of non-supersymmetric orbifolds
essentially consists of turning on the twisted sectors of the theory
that correspond to relevant (or marginal) deformations, and studying
the decay of the orbifolds under these \cite{hkmm}. In the NSR formalism for
closed string theory on generic orbifolds, the tachyonic deformations 
correspond to the orbifold twisted chiral operators with the (total) 
R-charge less than 
unity, while marginal deformations will have the total R-charge 
adding to one. Turning on these deformations, one can study the
RG flow of the resulting theory, and these nicely describe the decay
of non-supersymmetric orbifolds \cite{hkmm}. This is in turn closely
related to the toric geometry of non-supersymmetric orbifolds. 
Indeed, It was shown in \cite{mm} that the generators of the chiral
ring for these orbifolds are in one to one correspondence with the
minimal resolution curves of the singularities of $\BC^2/\BZ_n$, and
the R-charges of these are closely related to the integers appearing
in the Hirzebruch-Jung continued fraction that provides the intersection
numbers of the $\BC\BP^1$s involved in the resolution of the
singularity. 

Specifically, the toric geometry of the closed string SCFT probing
generic orbifolds of $\BC^2$ or $\BC^3$ can be studied by considering
the orbifold twisted sectors, that are relevant (or marginal).  
The toric data for the resolution of these orbifolds can be obtained
most simply by adding certain fractional points corresponding to 
the orbifold action (in a $\BZ^{\oplus r}$ lattice for an orbifold of 
$\BC^r$) and then restoring integrality in the lattice (see, for
eg. \cite{reid},\cite{ag},\cite{greene}). Let us see if we can 
substantiate this. Consider the orbifold $\BC^2/\BZ_{5(2)}$. 
In this case, there are four twisted sectors, and the relevant
deformations correspond to those with R-charges 
\cite{hkmm}
\begin{equation}
\left(\frac{1}{5},\frac{2}{5}\right),~~~  
\left(\frac{3}{5},\frac{1}{5}\right),  
\end{equation}
and these are the generators of the chiral ring of the orbifold SCFT.
The toric data for this orbifold is obtained by considering the
integral $\BZ^{\oplus 2}$ lattice generated by the vectors 
${\vec e_1}, {\vec e_2} = \left(1,0 \right), 
\left(0,1\right)$, but now with the fractional points 
$\left(\frac{1}{5},\frac{2}{5}\right), \left(\frac{3}{5},\frac{1}{5}\right)$,
which we call ${\vec e_3}, {\vec e_4}$ respectively. 
In order to restore integrality in the augmented lattice,
we now express the vectors in terms of ${\vec e_3}$, ${\vec e_2}$,
which we now label as $\left(1,0\right), \left(0,1\right)$ 
respectively. Doing this, we get the toric data as
\be
{\cal T} = 
\pmatrix{1&0&-1&-3\cr 0&1&3&5}
\ee
This is easily seen to be the toric data for the orbifold 
$\BC^2/\BZ_{5(2)}$, corresponding to the continued fraction $\frac{5}{2}$. 
\footnote{In general for more complicated orbifold theories, 
there might be more than one way to restore 
integrality in a lattice. These are essentially row-equivalent, 
but as a curiosity, we note that the number of these possible operations 
might be related to the multiplicities of the GLSM fields found via the 
open string picture in \cite{tapo2}.} 

A similar analysis can be done for orbifolds of the form 
$\BC^3/\BZ_m\times \BZ_n$, both for the space-time supersymmetric
and non-supersymmetric cases. In this case, there will be two 
generators of the orbifold action, which we will choose to be
\begin{eqnarray}
g_1 : \left(Z^1,Z^2,Z^3\right) \to 
\left(\omega_1 Z^1, \omega_1^p Z^2,Z^3\right)
\nonumber\\
g_2 : \left(Z^1,Z^2,Z^3\right) \to
\left(\omega_2 Z^1, Z^2, \omega_2^q Z^3\right)
\end{eqnarray}
where $\omega_1 = e^{\frac{2\pi i}{m}}, \omega_2 = e^{\frac{2\pi i}{n}}$,
and $p$ and $q$ are integers. \footnote{We will assume $p$ and $q$ to
be positive in what follows.} Again, as a concrete example, let us 
consider the space-time supersymmetric orbifold $\BC^3/\BZ_3 \times \BZ_3$,
\cite{bglp},\cite{fhh} with the orbifolding action being
\begin{eqnarray}
g1 : \left(Z^1,Z^2,Z^3\right) \to \left(\omega_1 Z^1, \omega_1^2 Z^2,
Z^3\right) \nonumber\\
g_2 : \left(Z^1,Z^2,Z^3\right) \to
\left(\omega_2 Z^1, Z^2, \omega_2^2 Z^3\right)
\end{eqnarray}
where now $\omega = e^{\frac{2\pi i}{3}}$. In the brane probe picture,
one considers the regular representations of the group $\BZ_3
\times \BZ_3$ which acts non trivially on the space-time coordinates
as well as the Chan-Paton factors. Equivalently, in the closed string
description, we consider the $\BZ^{\oplus 3}$ lattice, generated
by ${\vec e_1} = \left(1,0,0\right)$, ${\vec e_2} = \left(0,1,0\right)$,
${\vec e_3} = \left(0,0,1\right)$, but now including the following 
seven fractional points ${\vec e_4}, \cdots,{\vec e_{10}}$ which correspond
to the seven surviving (marginal) sectors in the theory, in our lattice :
\begin{eqnarray}
&~&\left(\frac{1}{3},\frac{2}{3},0\right),
~\left(\frac{2}{3},\frac{1}{3},0\right),
~\left(\frac{1}{3},0,\frac{2}{3}\right),
~\left(\frac{2}{3},0,\frac{1}{3}\right),\nonumber\\
&~&\left(0,\frac{1}{3},\frac{2}{3}\right),
~\left(0,\frac{2}{3},\frac{1}{3}\right),
~\left(\frac{1}{3},\frac{1}{3},\frac{1}{3}\right)
\end{eqnarray}
where in the second line, we have included the fractional points 
corresponding to the action by $g1.g2$ as well.  
In this case there are various possibilities of restoring 
integrality in our lattice. All these are of course related by
$SL\left(3,\BR\right)$ transformations. In particular, if we 
rewrite the vectors in terms of ${\vec e_8}$, ${\vec e_6}$,
${\vec e_3}$, which we now label by $\left(1,0,0\right)$,
$\left(0,1,0\right)$, $\left(0,0,1\right)$ respectively, we obtain
the toric data
\begin{equation}
{\cal T} = \pmatrix{1&0&0&2&-1&2&3&1&-2&4\cr
0&1&0&2&0&1&2&1&0&3\cr 0&0&1&-3&2&-2&-4&-1&3&-6}
\end{equation}
After a few row operations, this can be recognised as the 
toric data for the resolution of the orbifold 
$\BC^3/\BZ_3\times\BZ_3$. 

Similar analyses can be carried out for product orbifolds that have
tachyons in some of the twisted sectors. Consider, for example, the
non-cyclic orbifold $\BC^3/\BZ_5 \times \BZ_5$, with the orbifold
action now being given by
\begin{eqnarray}
g_1 : \left(Z^1,Z^2,Z^3\right) \to 
\left(\omega Z^1, \omega^2 Z^2,Z^3\right)
\nonumber\\
g_2 : \left(Z^1,Z^2,Z^3\right) \to
\left(Z^1, \omega Z^2, \omega^2 Z^3\right)
\label{z5z5a}
\end{eqnarray}
where now $\omega=e^{\frac{2\pi i}{5}}$. The orbifolding action will
now introduce tachyons in the closed string spectrum, corresponding
to twisted sectors that are relevant. In this case, the toric data
can be found from the set of points (apart from the generators of
the $\BZ^{\oplus 3}$ lattice :
\begin{equation}
\left(\frac{1}{5},\frac{2}{5},0\right),
~\left(\frac{3}{5},\frac{1}{5},0\right),
~\left(0,\frac{1}{5},\frac{2}{5}\right),
\left(0,\frac{3}{5},\frac{1}{5}\right),
~\left(\frac{1}{5},0,\frac{1}{5}\right)
\label{z5z5b}
\end{equation}
Where all the twisted sectors are tachyonic. Again, we can restore 
integrality in the augmented lattice in the 
same way as before, and the toric data corresponding to this orbifold
is given by 
\begin{equation}
{\cal T} = \pmatrix{1&0&0&-1&-1&-2&3&-3\cr
0&1&0&1&2&1&-1&1\cr0&0&1&1&0&3&0&5}
\label{z5z5c}
\end{equation}
A similar analysis can be carried out for four fold orbifolds as well.

Once we obtain the toric data for a given orbifold, it is easy to
construct the charges of the GLSM fields that describes (in a classical 
limit) the orbifold singularity. In particular, the GLSM charge
matrix is given by the kernel of the toric data. Consider, for eg. 
the simplest non-cyclic orbifold, $\BC^3/\BZ_2\times \BZ_2$, whose
toric data is given by \cite{greene}
\begin{equation}
{\cal T} = \pmatrix{1&0&0&0&2&1\cr 0&1&0&-1&1&1\cr 0&0&1&2&-2&-1}
\end{equation}
The GLSM charge matrix is given by
\begin{equation}
Q = \pmatrix{1&0&0&-1&-1&1\cr 0&1&0&-1&1&-1\cr 0&0&1&1&-1&-1}
\end{equation}
Here we have three marginal deformations in the world sheet SCFT.
\footnote{The charge matrix of the GLSM can be written in various
bases. We will choose the basis that makes the resolution of the 
singularity clearly visible.} We can now look at the partial 
resolutions of this singularity \cite{nonsph}, by resolving
certain points in the toric diagram. In this example, if we consider
the toric data with two of the marginal sectors turned on, we 
obtain the GLSM charge matrix 
\begin{equation}
Q = \pmatrix{1&1&0&-2&0\cr 0&1&1&0&-2}
\end{equation}
which clearly specifies the action of the orbifolding group on $\BC^3$. 

The same analysis can be carried over in the case of non-supersymmetric
orbifolds as well. The GLSM charge matrix can be written down in a 
similar fashion as above, and using this, we can carry out an analysis
of the phases of the orbifold theory. Note that in general, we will have
GLSMs charged under multiple $U(1)$ gauge groups. For simplicity, 
we can, however, choose turn on a subset of the deformations (relevant or
marginal). Consider for eg. the non-supersymmetric orbifold 
$\BC^3/\BZ_5\times\BZ_5$ with the orbifolding action, the 
twisted sector R-charges and the toric data given by (\ref{z5z5a}), 
(\ref{z5z5b}) and (\ref{z5z5c}) respectively. Turning on two of the
twisted sectors with R-charges $\left(\frac{1}{5},\frac{2}{5},0\right)$
and $\left(0,\frac{1}{5},\frac{2}{5}\right)$, the GLSM charge matrix
is given by
\begin{equation}
Q = \pmatrix{1&2&0&-5&0\cr0&1&2&0&-5}
\end{equation}
In the next section, we will elaborate on this, and study in details the
structure of the GLSM with arbitrary charge matrices. This will enable
us to write down the classical metrics seen by the GLSM in certain 
limits, and will help us to analyse the phases of the models.  

Before we end this section, a few comments on GSO projections of the
string theories under consideration is in order. In \cite{mnp},
it was shown that in order for the string theory to admit a Type II
GSO projection in the orbifold $\BC^3/\BZ_{n(k_1,k_2,k_3)}$, 
we should have $\sum_i=1^3 k_i = {\mbox{even}}$. The GSO projection
acts nontrivially on the twisted sectors and the necessary condition
that these be preserved under a Type II GSO projection as well is
that the integer part of the R-charges of these twisted sectors 
should be even. In the GLSM, this amounts to demanding that
the sum of the $U(1)$ charges for each gauge group be even, in 
order to admit a Type II GSO projection \cite{narayan}. 
In the case of type 0 string theories, on the other hand, the 
presence of the bulk tachyon might complicate the dynamics of the 
decay. However, we assume in these cases that the delocalised bulk 
tachyon is tuned such that it does not affect the dynamics of the RG 
flow, which is driven solely by the twisted sector tachyons. Since 
our analysis of the next section is completely general, we will not 
mention this point explicitly in future, and the GSO projection will 
be understood from the context.

\section{Sigma Model Metrics for Multi $U(1)$ GLSMs}

In this section, we study in details the multi parameter GLSMs, 
generalising the results of \cite{ts1},
\cite{mt}. This will be useful for us when we consider the
phases of generic orbifolds of the form $\BC^3/\BZ_n\times\BZ_m$. 
We will first start by reviewing briefly the relevant details of
Wittens's GLSM \cite{wittenphases}. We then construct the bosonic
GLSM Lagrangian for arbitrary charges, and proceed to evaluate
the sigma model metrics for the same. 

\subsection{Single Parameter GLSMs}

Witten's Gauged Linear Sigma model \cite{wittenphases}
provides an effective tool for studying closed string tachyon 
condensation on orbifolds \cite{vafa}. The action for the GLSM,
with an Abelian gauge group $U(1)^s$ is

\be
S = \int d^2z d^4\theta\sum_i {\bar\Phi_i}\Phi_i
-\sum_a\frac{1}{4e_a^2}\int d^2z d^4\theta{\bar\Sigma_a}\Sigma_a 
+ {\mbox{Re}}\left[it\int d^2z d^2{\tilde\theta}\Sigma\right]
\label{glsmaction}
\ee
where the $\Phi_i$ are chiral superfields, $\Sigma_a$ are twisted
chiral superfields ($a = 1 \cdots s$), $t = ir + \frac{\theta}{2\pi}$ 
is a complexified parameter involving the Fayet-Iliopoulos parameter 
$r$ and the two dimensional $\theta$ angle, and we are considering
a theory without any superpotential.  

The $e^2 \to \infty$ limit of the GLSM is the non-linear sigma model 
(NLSM) limit, and in this limit, the gauge fields appearing in 
(\ref{glsmaction}) appear as Lagrange multipliers. It is then 
possible to solve the D-term constraint in the classical limit 
$|r| \to \infty$, to read off the sigma model metric corresponding
to the GLSM \cite{mt},\cite{ts1}. 
It will be enough for our purpose to focus on the bosonic 
part of the action in (\ref{glsmaction}), given by
\be
S = -\int d^2z D_{\mu}{\bar \phi_i}D^{\mu}\phi_i
\label{glsmbosonic}
\ee
and study this action, using the D-term constraints,
\be
\sum_a Q_i^a{\bar \phi_i}\phi_i = 0
\label{dterm}
\ee
where $\phi_i$ are the bosonic components of $\Phi_i$ and 
$Q_i^a$ denote the charge of the $\phi_i$ with respect to the
$a$ th $U(1)$.    

Orbifolds of the type $\BC^r/\Gamma$, with $r=1,2,3$ can be
effectively described by a single parameter GLSM, with the number
of gauge groups being dictated by the nature of the singularity.
In this subsection, we will consider the single parameter GLSM, 
which describes closed string tachyon condensation in 
$\BC/\BZ_n$ (with $n$ being odd for localised tachyons \cite{aps}),
and can also be used to describe condensation of a single tachyon
in orbifolds of $\BC^2$ and $\BC^3$. Let us start by reviewing the
process of closed string tachyon condensation in $\BC/\BZ_n$.

In this case, the GLSM consists of two fields, charged under a 
single $U(1)$ gauge group, with charges $\left(1, -n\right)$, and
satisfies the D-term constraint given by
\be
|\phi_1|^2 - n|\phi_{2}|^2 + r = 0
\label{dtermsingleu1}
\ee
where $r$ is the Fayet-Iliopoulos parameter of the theory. 
The D-term constraint is solved by
\be
\phi_1 = \rho_1 e^{i\theta_1} ~~~~ \phi_2= \rho_2 e^{i\theta_2}
= \sqrt{\frac{\rho_1^2 + r}{n}}e^{i\theta_2}
\label{solonepar}
\ee
The bosonic Lagrangian (\ref{glsmbosonic}) can be used to solve 
classically for the gauge field, with the solution being, in this
case, 
\be
V_{\mu} = \frac{\sum_i Q_i \left({\bar\phi_i}\partial_{\mu}\phi_i
- \phi_i\partial_{\mu}{\bar\phi_i}\right)}{2i\sum Q_i^2|\phi_i|^2}
\label{vonepar}
\ee 
Upon substituting (\ref{solonepar}) in (\ref{vonepar}), and putting
it back in the action (\ref{glsmbosonic}), we obtain the Lagrangian 
\be
L = \left(\partial_{\mu}\rho_1\right)^2 + 
\left(\partial_{\mu}\rho_2\right)^2 + 
\frac{\rho_1^2 \rho_2^2 \left(Q_1 d\theta_2 - Q_2 d\theta_1\right)^2}
{\left(Q_1^2\rho_1^2 + Q_2^2\rho_2^2\right)} 
\label{lagonepar2}
\ee
In the limit when $r \to \infty$, substituting for $\rho_1$ and
$\rho_n$ from (\ref{solonepar}) in (\ref{lagonepar2}, 
we recover the sigma model metric of the cone, $\BC/\BZ_n$,
\be
ds^2 = d\rho_1^2 + \frac{\rho_1^2}{n^2}d\theta^2
\ee
where $\theta = \left(nd\theta_2 - d\theta_1\right)$ 
is the gauge invariant angle. This geometry can also be seen from the 
D-term equation (\ref{dtermsingleu1}). For large $r$, $\phi_{n}$ 
acquires a large vev, and the gauge group $U(1)$ is broken to $\BZ_n$. 
A similar calculation \cite{mt} with $r \to -\infty$ shows that the space is
now flat (after tachyon condensation), in agreement with the predictions
of \cite{aps}. 

In the same way, single parameter GLSMs with more fields can be considered.
It is easy to carry out the analysis of the sigma model metric in 
exactly the same way above \cite{ts1} and we will write the result for 
the Lagrangian for a GLSM with $m$ fields $\phi_i, i=1,\cdots m$ 
with $U(1)$ charges $Q_i,~~i=1,\cdots m$ :
\be
L = \left(\partial_{\mu}\rho_1\right)^2 + 
\left(\partial_{\mu}\rho_2\right)^2 + \cdots +
\left(\partial_{\mu}\rho_m\right)^2 +  
\frac{\sum_{i <  j}\rho_i^2\rho_j^2\left(Q_id\theta_j 
- Q_jd\theta_i\right)^2}{\sum_i Q_i^2\rho_i^2}
\label{lagonepargen}
\ee
The above formula gives the single parameter GLSM Lagrangian for 
the singularity $\BC^{m-1}/\BZ_n$ and is the main result of this
subsection.
 
The sigma model metrics corresponding to the various limits of the 
Fayet-Iliopoulos parameter can again be obtained by solving the D-term
constraints. As an example, the supersymmetric orbifold 
$\BC^3/\BZ_{3(-1)}$, which is described by
\be
Q_i = \left(1,1,1,-3\right),~~~~~|\phi_1|^2 + |\phi_2|^2
+ |\phi_3|^2 -3|\phi_4|^2 + r = 0 
\ee
can be seen, in the limit $r \to \infty$ to have the sigma model metric
\be
ds^2 = d\rho_1^2 + d\rho_2^2 + d\rho_3^2 +
\frac{\rho_1^2}{9}d{\tilde\theta_1}^2 + \frac{\rho_2^2}{9}d{\tilde\theta_2}^2 +
\frac{\rho_3^2}{9}d{\tilde\theta_3}^2 
\ee
where ${\tilde\theta_1} = \left(3\theta_1 + \theta_4\right)$,
${\tilde\theta_2} = \left(3\theta_2 + \theta_4\right)$,
${\tilde\theta_3} = \left(3\theta_3 + \theta_4\right)$ are the gauge
invariant combinations of the original phases appearing in the solutions
for the $\phi_i$s. The sigma model metric in the limit $r \to -\infty$ can 
also be read off from this Lagrangian, and gives (three copies of) 
the flat space metric. This corresponds to ``blowing up'' the singularity
using marginal operators in the CFT, as in \cite{hkmm}. Away from the
classical limits (i.e when $r$ is small), the sigma model metrics 
receive quantum corrections \cite{mt}. For our purposes, it will be
enough to consider only the $|r| \to \infty$ limits in the above
Lagrangian.   

For tachyonic orbifolds, the general procedure outlined above can be used to 
study the decays of the singular theory, under tachyon condensation. 
For eg. the tachyonic orbifold $\BC^2/\BZ_{n(k)}, 1+k \neq 0 {\mbox{mod}}n$
can be modelled by the single parameter GLSM with three fields, with
the $U(1)$ charges being $Q_i = \left(1,k,-n\right)$. In this case,
for the limit $r \to \infty$, the D-term constraint 
\be
|\phi_1|^2 + k|\phi_2|^2 - n|\phi_3|^2 + r = 0
\label{dtermcons3}
\ee
is solved by setting
\be
\phi_1 = \rho_1e^{i\theta_1}, \phi_2 = \rho_2e^{i\theta_2},
\phi_3 = \rho_3 e^{i\theta_3}=\sqrt{\frac{r+\rho_1^2+k\rho_2^2}{n}}
\label{dterm3}
\ee
In the limit when $r \to \infty$, substituting (\ref{dterm3}) in
(\ref{lagonepargen}), we recover the metric \cite{ts1}
\be
ds^2 = d\rho_1^2 + d\rho_2^2 + \frac{\rho_1^2}{n^2}d{\tilde\theta_1}^2
+ \frac{\rho_2^2}{(n/k)^2}d{\tilde\theta_2}^2 
\ee
where the gauge invariant angles are now ${\tilde\theta_1}=
\left(\theta_3 + n\theta_1\right), {\tilde\theta_2}=
\left(\theta_3 + \frac{n}{k}\theta_1\right)$. The geometry 
is most easily visualised by making the gauge choice $\theta_n=0$
which fixes the metric to be
\be
ds^2 = d\rho_1^2 + d\rho_2^2 + \rho_1^2d\theta_1^2 + \rho_2^2d\theta_2^2
\ee
but with the simultaneous identifications
\be
\theta_1 \simeq \theta_1 + \frac{2\pi}{n}, ~~
\theta_2 \simeq \theta_2 + \frac{2\pi k}{n}
\ee
The $r \to -\infty$ limit can be worked out in an entirely analogous
way. In this case the D-term constraint (\ref{dtermcons3}) shows that
$\phi_1$ and $\phi_2$ cannot be simultaneously zero, and choosing $\phi_k$ 
to be very large, the D-term constraint can be solved  by setting
\be
\phi_1=\rho_1e^{i\theta_1}, \phi_2 = \sqrt{\frac{\rho_3^2 - \rho_1^2 -r}
{k}}, \phi_3 = \rho_3e^{i\theta_3}
\ee
Putting these solutions in the Lagrangian (\ref{lagonepargen}) yields
the metric
\be
ds^2 = d\rho_1^2 + d\rho_3^2 + \frac{\rho_1^2}{k^2}d{\tilde\theta_1}^2
+ \frac{\rho_3^2}{\left(k/\left(jk-n\right)\right)^2}d{\tilde\theta_2}^2
\ee
where the gauge invariant angles are ${\tilde\theta_1} = 
\left(\theta_k - k\theta_1\right), {\tilde\theta_2} = 
\left(\theta_k - \frac{k}{\left(2k - n\right)}\right)$, and 
according to our convention of the last section, the D-term
constraint (\ref{dtermcons3}) necessitates the redefinition of the 
charge of $\phi_3$ as $n \to \left(n-jk\right)$, where $j$ is the 
smallest integer that makes $\left(jk - n\right)$ positive.  
Likewise, when $\phi_1$ is made very large, it can be checked that 
we recover flat space. This describes the decay of the orbifold
$\BC^2/\BZ_{n(k)}$ via turning on a single $U(1)$. The resulting
space(s) can of course be singular themselves, in which case it is
necessary to turn on a second $U(1)$ and follow its flow, and so on.  
The details of this can be found in \cite{ts1}. \footnote{Note
that our GLSM analysis of $\BC^2/\BZ_{n(k)}$ essentially 
describes the condensation of the tachyon of the first twisted sector 
of the closed string CFT. In order to study the condensation of the 
tachyon in the $j$th twisted sector, we need to modify the GLSM to have 
the charges $Q_i = \left(j, jk, -n\right)$ and the sigma model metrics
can be calculated as usual.}

\subsection{Two Parameter GLSMs}

We now move over to the description of two parameter GLSMs. The 
calculations are straightforward but lengthy, and we will simply 
present the final results. In this case, it is easy to check that
varying the bosonic action (\ref{glsmbosonic}), we get the following
equations for the gauge fields 
\be
\sum_iQ_i^a\sum_b Q_i^bV_{\mu}^b= Q_i^a\sum_i{\mbox{Im}}\left(
{\bar \phi_i}\partial_{\mu}\phi_i\right)
\label{meq}
\ee
where $a,b = 1, 2$, and
the sum over $i$ goes over the chiral fields. Eq. (\ref{meq}) 
gives a set of simultaneous equations 
and can be readily solved, and it is seen that in this case, 
integrating out $V_{\mu}^a$, i.e solving for $V_{\mu}^a$ from 
(\ref{meq}), and substituting in the Lagrangian of (\ref{glsmbosonic}) 
yields, after a somewhat lengthy computation 
\be
L = L_1 + L_2
\label{lagtwopargen}
\ee
where
\be
L_1 = \sum_i\left(\partial_{\mu}\rho_i\right)^2
\label{l1}
\ee
\begin{equation}
L_2 = \frac{\sum_{[i,j,k]}\left[\rho_i\rho_j\rho_k
d\theta_i\left(Q_j^bQ_k^a - Q_k^bQ_j^a\right)\right]^2}
{\sum_{i<j}\rho_i^2\rho_j^2\left(Q_i^bQ_j^a - Q_j^bQ_i^a\right)^2}
\label{l2}
\end{equation}
Where we have written $\phi_i = \rho_ie^{i\theta_i}$, 
\footnote{From now on, we will not explicitly solve the D-term
constraint. Rather we write the GLSM fields $\phi_i=
\rho_ie^{i\theta_i}$ and write the Lagrangian in terms of 
$\rho_i$s and $\theta_i$s. This will help us to find a 
general expression in which we can make any field acquire an
arbitrarily large vev, so that it is integrated out} and the
symbol $[i,j,k]$ in the summation in the numerator of $L_2$ denotes 
{\it cyclic} combinations of the variables. 
The equations (\ref{lagtwopargen}), (\ref{l1}) and (\ref{l2}) are 
the main result of this subsection.

The details of the calculation are unimportant, but as a check, let
us consider the single parameter GLSM with three fields having charges
\be
Q_i = \left(1, n_2, -n_3\right)
\ee
which, in the limit of the Fayet-Iliopoulos parameter going to 
infinity, describes the unresolved orbifold $\BC^2\BZ_{n_3(n_2)}$. 
The metric calculated from (\ref{lagonepargen}) yields
\be
ds^2 = d\rho_1^2 + d\rho_2^2 + \frac{\rho_1^2}{n_3^2}d{\tilde\theta_1}^2
+ \frac{\rho_2^2}{(n_3/n_2)^2}d{\tilde\theta_2}^2
\label{check1}
\ee
where ${\tilde\theta_1} = \theta_3 + n_3\theta_1,~~
{\tilde\theta_2} = \theta_3 + \frac{n_3}{n_2}\theta_2$ are the
gauge invariant angles, with $\theta_1$ and $\theta_2$ appearing in
the original solutions of $\phi_1$ and $\phi_2$. The same metric
can be calculated by using a two parameter model, with a second 
$U(1)$ corresponding to the $j$th twisted sector of the CFT being
turned on. Using the charges 
\be
Q_i = \pmatrix{n_1&n_2&n_3&0\cr jn_1& jn_2 & 0 & n_3}
\ee
which indeed describes the completely unresolved orbifold 
$\BC^2/\BZ_{n_3(n_2)}$ (as can be seen by simply writing down the
two D-term constraints), in (\ref{l1}) and (\ref{l2}), we obtain
the sigma model metric
\be
ds^2 = d\rho_1^2 + d\rho_2^2 + 
\frac{\rho_1^2}{n_3^2}d{\tilde\theta_1}'^2 
+ \frac{\rho_2^2}{(n_3/n_2)^2}d{\tilde\theta_2}'^2
\ee
where now $\theta_1'$ and $\theta_2'$ are given by the expressions 
\be
\theta_1' = \left(j\theta_4 + \theta_3\right) + n_3\theta_1,~~~
\theta_2' = \left(j\theta_4 + \theta_3\right) + \frac{n_3}{n_2}\theta_2
\ee
This is seen to match with (\ref{check1}) by a trivial redefinition
of the angles. 

Note that in our calculations leading to the Lagrangians in 
(\ref{lagonepargen}) and (\ref{lagtwopargen}), we have used the homogeneous
coordinates of the GLSM. It is possible to perform the analysis 
presented here by using symplectic coordinates, where it turns out
that the essential information about the metric is given by the 
Hessian of the {\it symplectic potential} \cite{sureshnotes}, by
generalising an approach due to Guillemin \cite{guillemin} 
(see also \cite{koushik}). However, we wish to assert that our
calculation above presents the Lagrangian entirely in terms of the
$U(1)$ charges of the GLSM, which is the kernel of the toric data
of the orbifold singularity, and is completely general in that aspect.
We will therefore use the above analysis in a general manner while
dealing with generic multi parameter GLSMs.  

The analysis presented above can be used to study non-cyclic 
singularities of the form $\BC^3/\BZ_m\times\BZ_n$, which cannot
be described by a single parameter GLSM. Before we embark on the full
details in the next section, let us end this section by presenting
a couple of results for the metrics of completely unresolved 
supersymmetric orbifolds. Our first example is the orbifold
$\BC^3/\BZ_{4(1,1,2)}$. This can be described by the two-parameter
GLSM with charges
\be
Q_i = \pmatrix{1&1&2&-4&0\cr 2&2&0&0&-4}
\ee
The metric, in this case, is given by
\be
ds^2 = d\rho_1^2 + d\rho_2^2 + d\rho_3^2 +
\frac{\rho_1^2}{16}d{\tilde\theta_1}^2 +
\frac{\rho_2^2}{16}d{\tilde\theta_2}^2 +
\frac{\rho_3^2}{4}d{\tilde\theta_3}^2
\ee
with 
\be
{\tilde\theta_1} = \left(\theta_4 + 4\theta_1 + 2\theta_5\right),~~
{\tilde\theta_2} = \left(\theta_4 + 4\theta_2 + 2\theta_5\right),~~
{\tilde\theta_3} = \left(\theta_4 + 2\theta_3\right)
\ee
Our last example in this section is the supersymmetric orbifold
$\BC^3/\BZ_2\times\BZ_2$. This can be described by the $U(1)^2$
GLSM with five fields, having, in a certain basis, the charges
\footnote{In the notation of the last section, we have turned on
two of the three twisted sectors, corresponding to the charges 
$\left(\frac{1}{2}, \frac{1}{2}, 0\right)$ and 
$\left(\frac{1}{2}, 0, \frac{1}{2}\right)$}
\be
Q_i = \pmatrix{1&1&0&-2&0\cr 0&1&1&0&-2}
\ee
The corresponding toric data being
\be
{\cal T} = \pmatrix{1&0&0&0&2\cr 0&1&0&-1&1\cr 0&0&1&2&-2}
\ee
The sigma model metric, calculated from (\ref{lagtwopargen}) yields
\be
ds^2 = d\rho_1^2 + d\rho_2^2 + d\rho_3^2 + \frac{\rho_1^2}{4}
d{\tilde\theta_1}^2 + \frac{\rho_2^2}{4}d{\tilde\theta_2}^2
+ \frac{\rho_3^2}{4}d{\tilde\theta_3}^2 
\ee
where we have defined
\be
{\tilde\theta_1} = \left(\theta_4 + 2\theta_1 \right),~~
{\tilde\theta_2} = \left(\theta_5 + 2\theta_3 \right),~~
{\tilde\theta_3} = \left(\theta_4 + \theta_5 + 2\theta_2\right)
\ee

\subsection{General $r$ parameter GLSMs}

We are now ready to write down the result for the Lagrangian for the
general $r$ parameter GLSM. Once again, the details are unimportant (as
much as they are lengthy), and we simply present the result here. 
We find that in the general case, the $r$ parameter GLSM Lagrangian 
can be written as
\be
L = L_1 + L_2
\label{lagrpargen}
\ee
where now
\be
L_1 = \sum_i\left(\partial_{\mu}\rho_i\right)^2
\label{mostgenl1}
\ee
\begin{equation}
L_2 = \frac{\sum_{[j_1,j_2,\cdots,j_{r+1}]}\left[
\rho_{j_1}\rho_{j_2}\cdots\rho_{j_{r+1}}\partial_{\mu}
\left(\theta_{j_1}K_{j_2,\cdots,j_r}\right)\right]^2}
{\sum_{j_1<j_2<\cdots j_r}\rho_{j_1}^2\rho_{j_2}^2
\cdots\rho_{j_r}^2\left[\Delta\left(j_1,j_2,\cdots,j_r\right)\right]^2}
\label{mostgenl2}
\end{equation}
where $i$ and $j_1,j_2,\cdots, j_{r+1}$ go from $1,2,\cdots,r+m$ for a 
$\BC^m/\Gamma$ orbifold, $K_{j_2,\cdots,j_r}$ is the $j_1$th component of the 
Kernel of the matrix formed by the charges of the 
$j_{r+1}$ vectors in the numerator of $L_2$ (and hence depends on
$j_2,\cdots j_{r+1}$), and 
$\Delta\left(j_1,j_2,\cdots,j_r\right)$ is the determinant of the 
matrix formed by the charge vectors $\rho_{j_1}, \rho_{j_2}
\cdots\rho_{j_r}$ under the $r$ $U(1)$s. Also, as before, the
notation $[j_1,j_2,\cdots,j_{r+1}]$ indicates a cyclic combination
of the variables. \footnote{Operationally, the process of giving
large vevs to certain fields so as to integrate them out is 
similar to resolving certain points in the toric diagram 
corresponding to the singularity. It may look like there is a
potential ambiguity in the definition of the charges when certain
fields are resolved, since the squares of the charges appear 
in the denominator of (\ref{mostgenl2}). However, this can
be fixed by appealing to the consistency of the toric data at
each step. We will have more to say about this towards the end of
this section.}    

The advantage of the formula (\ref{lagrpargen}) above is that it is 
completely general
and we have written it {\it entirely} in terms of the toric data of the
orbifold. One can check that in the special cases of the 
one and the two parameter GLSMs, eq. (\ref{lagrpargen}) reduces to the
eqs. (\ref{lagonepargen}) and (\ref{lagtwopargen}) respectively. 
Eq. (\ref{lagrpargen}) is the main result of this subsection.

The formulation above can be applied to study the dynamics of 
D-branes in GLSMs, by using the boundary GLSM approach of 
\cite{hiqv},\cite{tapo3}. In particular, we may hope that by using
the open string world sheet description of the GLSM Lagrangian
that we have discussed, we might be able to derive general D-brane 
boundary conditions, for multi parameter GLSMs, and study branes
in various phases of these. We will comment on this towards the
end of the paper.  

Having written down the GLSM Lagrangian in its most general form
entirely in terms of the toric data of the orbifold, we can now use
this formalism to gain knowledge about the phases of the GLSMs for
any given orbifold. Essentially these phases will be obtained from
the Lagrangian by making some fields in the GLSM very large. Since
we have the Lagrangian in the most general form, we do not need to
explicitly solve the D-term constraints \footnote{Excepting that in
eq. (\ref{mostgenl1}) we need to set the terms corresponding to the
``large fields'' to zero, since they are implicitly solved by constraints 
of the form (\ref{solonepar}) and (\ref{dterm3}), and as can be easily
seen, drop out of the calculation} but can work directly with the 
fields.  

As a check on our formula, note that for $r=2$, it reduces to 
eq. (\ref{lagtwopargen}). Consider a generic $\BC^3/\Gamma$ 
singularity, with five fields and a $U(1)^2$ charge matrix being given by
\begin{equation}
Q = \pmatrix{Q_1^1&Q_2^1&Q_3^1&Q_4^1&Q_5^1\cr 
Q_1^1&Q_2^1&Q_3^1&Q_4^1&Q_5^1}
\end{equation}
where the subscripts on the charges label the fields and the superscripts
label the $U(1)$. Now, in the Lagrangian of eq. (\ref{lagtwopargen}),
we take the classical limit $\rho_1 \to \infty$. It is easy to see that
in this limit, the Lagrangian simplifies to one of four fields, charged
under a single $U(1)$, with the charges now being 
\begin{equation}
Q = \left(Q_1^1Q_2^2 - Q_1^2Q_2^1,~Q_1^1Q_3^2 - Q_1^2Q_3^1,~
Q_1^1Q_4^2 - Q_1^2Q_4^1,~Q_1^1Q_5^2 - Q_1^2Q_5^1\right)
\label{chargetwopar}
\end{equation}
These are the relations that define the new D-term constraints in
terms of the single $U(1)$ \footnote{Note that we have to ignore
any common numerical factor that appears in the expression for the
charges in (\ref{chargetwopar}). In our Lagrangian of 
eq. (\ref{lagrpargen}), these numerical factors will cancel out 
automatically in (\ref{mostgenl2}) and in (\ref{mostgenl1}), we
will assume that the fields that have been set very large do not
contribute because of the implicit nature of the solution for the D-term
constraints}. Operationally, this is equivalent to
removing one point from the toric data given by the original charge 
matrix, and taking the kernel of the new toric data so obtained to
get the new charge matrix. For the case of generic $U(1)^r$ GLSMs,
general expressions for the charges of the fields in the reduced 
Lagrangian get algebraically complicated, and we will not 
present the results here. We have checked that in general, by taking
the classical limits in which certain fields are made very large
reduces the Lagrangian (\ref{lagrpargen}) to an appropriate 
Lagrangian with a lower rank gauge group.

Before we end this section, a few comments are in order. First of all,
consider the supersymmetric orbifolds of the form $\BC^3/\BZ_n$,
with the $U(1)$ charge matrix being given by
\be
Q=\left(Q_1, Q_2, Q_3, Q_4\right)
\ee
with $\sum Q_i = 0$. In the limit when the Fayet-Iliopoulos parameter 
is very large (so that one of the fields, say with charge $Q_4$ is 
always large and positive), we can consider the 
GLSM Lagrangian with one of the other fields set to zero, so that
our Lagrangian now represents an appropriate $\BC^2$ singularity. 
There are three of these, and when we add them up, we reproduce (apart 
from a trivial factor of $2$) the original GLSM Lagrangian in 
this limit. This is one of the results of \cite{cr}, termed
``champions meet'' in that paper. 

Secondly, note for the special case of product spaces, the GLSM 
Lagrangian of eq. (\ref{lagrpargen}) is separable. Consider, for example,
the surface ${\cal F}_0 = \BC\BP^1\times \BC\BP^1$. This is described
by the GLSM with four fields, and a $U(1)^2$ charge matrix
\be
Q = \pmatrix{1&1&0&0\cr 0&0&1&1}
\ee
The Lagrangian corresponding to this space can be seen to be a sum
of two $\BC\BP^1$ Lagrangians. 
 
Also, note that there might be a potential ambiguity in the sign of
the redefined charges, after giving large vevs to certain fields. 
This is due to the fact that the square of the charge is what appears
in the denominator of eq. (\ref{mostgenl2}) or eq. (\ref{l2}).
This is not a problem, if we note that by resolving fields in succession,
we finally reach a single parameter GLSM, which effectively describes
the same singularity. Operationally, the sign of the charge of the
remaining fields is obtained by first writing the fields that take
large vevs (in decreasing order, as in (\ref{lagrpargen})), and then
writing the remaining fields. The determinant in the denominator of
(\ref{lagrpargen}) changes sign appropriately, with the absolute
value of the charges remaining the same.  

Having discussed the generic $r$ parameter GLSMs, We are now ready 
to use the results of this section to analyse the phases
of generic multiparameter GLSMs.  
 
\section{Phases of Generic Multiparameter GLSMs}

In this section, we analyse the phases of generic multiparameter
GLSMs, using the formalism of the last section. We will not attempt
to deal with GLSMs with generic charges, since this is algebraically 
cumbersome beyond the two parameter case. Also, we leave a general 
treatment of these 
phases in lines with \cite{mn} for generic orbifolds of $\BC^3$, of the 
form $\BC^3/\BZ_m\times\BZ_n$ for the future. We will rather try 
to apply the results of the last section to some specific examples. 
Consider, for example, the orbifold
$\BC^3/\BZ_{13(1,2,5)}$ \cite{mn}. \footnote{This orbifold has
been studied in details in \cite{mn}. We will, however focus on 
this example in order for ease of comparison with the
methodology of that paper.} Closed string tachyon condensation can be 
studied in this Type 0 example. There are three twisted sector 
tachyons in the closed string spectrum, and the $U(1)^3$ charge matrix 
is given by
\begin{equation}
Q = \pmatrix{1&2&5&-13&0&0\cr 8&3&1&0&-13&0\cr 3&6&2&0&0&-13}
\label{threetachyons}
\end{equation}
This GLSM describes the orbifold with three tachyons, and 
in order to study the phases of this theory, we can write the 
GLSM Lagrangian for this charge configuration by specialising
to the case $r=3$ in (\ref{lagrpargen}). For ease of comparison, 
we will begin, however, by turning on two of the tachyons, and
the GLSM charge matrix for five fields with these two tachyons 
represented by the last two columns of the charge matrix
\begin{equation}
Q = \pmatrix{1&2&5&-13&0\cr8&3&1&0&-13}
\label{twotachyons}
\end{equation} 
The GLSM Lagrangian can be read off directly from eq.
(\ref{lagtwopargen}) or by specialising to the case of $r=2$
in eq. (\ref{lagrpargen}). The phase boundaries (in the space of
the Fayet-Iliopoulos parameters) can be read off from the 
columns of the charge matrix (\ref{twotachyons}) and the sigma
model metrics for the phases of the model can be calculated by
considering making two fields large in eq. (\ref{lagtwopargen}).
We will simply present the results for the sigma model metrics that
can be read off from the Lagrangian in the appropriate limits. In
the phases where the non-tachyonic fields are large, the 
classical sigma model metrics are
\begin{eqnarray}
ds_{12}^2 &=& d\rho_3^2 + d\rho_4^2 + d\rho_5^2 + 
\rho_3^2 d\left(\theta_1 - 3\theta_2 + \theta_3\right)^2 + \nonumber\\
&~&\rho_4^2 d\left(-3\theta_1 + 8\theta_2 + \theta_4\right)^2 +
\rho_3^2 d\left(2\theta_1 - \theta_2 + \theta_5\right)^2 +
\nonumber\\
ds_{13}^2 &=& d\rho_2^2 + d\rho_4^2 + d\rho_5^2 +
\frac{\rho_2^2}{9}d\left(\theta_1 + \theta_3 -3\theta_2\right)^2 +\nonumber\\
&~&\frac{\rho_4^2}{9}d\left(\theta_1 + \theta_3 -3\theta_4\right)^2 +
\frac{\rho_5^2}{9}d\left(\theta_1 + \theta_3 -3\theta_5\right)^2 
\nonumber\\
ds_{23}^2 &=& d\rho_1^2 + d\rho_4^2 + d\rho_5^2 +
\rho_1^2 d\left(-3\theta_2 + \theta_3 + \theta_1\right)^2+\nonumber\\
&~&\rho_4^2 d\left(-\theta_2 + 3\theta_3 + \theta_4\right)^2+
\rho_1^2 d\left(5\theta_2 - 2\theta_3 + \theta_5\right)^2+
\end{eqnarray}
where each of the five fields $\phi_i,~i=1,\cdots,5$ have been written
as $\phi_i = \rho_ie^{i\theta_i}$ and 
the subscripts label the fields that have been given large vevs, and
in the space of the Fayet-Iliopoulos parameters denote the region in 
which the metric is valid.  
It can be seen from above that whereas the first and third regions. 
represent flat space, the second is the unresolved phase of the 
supersymmetric GLSM with charges $Q=\left(1,1,1,-3\right)$, by 
an appropriate redefinition of the gauge invariant angles. In the 
same spirit, we can analyse the sigma model metrics for the other 
regions of moduli space. We summarise the results below 
\begin{eqnarray}
&~& ds_{14}^2 \sim \left(1,1,3,-8\right),~~
ds_{15}^2 \sim {\mbox{flat}},~~
ds_{23}^2 \sim {\mbox{flat}}, \nonumber\\
&~& ds_{24}^2 \sim \left(-1,1,-1,-3\right),~~
ds_{25}^2 \sim \left(1,1,1,-2\right),\nonumber\\
&~& ds_{34}^2 \sim {\mbox{flat}},~~
ds_{35}^2 \sim \left(1,1,3,-5\right),~~
ds_{45}^2 \sim \left(1,2,5, -13\right)
\label{various}
\end{eqnarray}  
In the above equation, we have written the metrics implicitly, for eg. when 
the fields $\phi_1, \phi_4 \gg 0$, the resulting metric is that of a
GLSM with charges $\left(1,1,3,-8\right)$ in the infrared limit (i.e
the unresolved phase of the latter).  
Eq. (\ref{various}) above gives us the behaviour of the classical metrics
in different regions of moduli space. The distinct {\it phases} of the
theory can be read off straightforwardly, by listing the 
massless fields that remain after certain fields have been given
large vevs. For eg., noting that in the two dimensional space of the 
Fayet-Iliopoulos parameters, the phase boundaries are vectors 
denoted by their charges in (\ref{twotachyons}) \cite{mn}
and for eg. the overlap region where $ds_{15}^2$, $ds_{25}^2$,
$ds_{35}^2$ are all valid gives the region in moduli space where
the tachyon field $\phi_4$ is massless, we see that this region is the
phase denoting partial resolution by the tachyon field $\phi_4$. 
A similar logic for the other metrics gives the phase diagram for
this model. Importantly, we can read off the sigma model metrics 
not only in the classical regions of all the phases, but other
generic points also, i.e whenever some fields in the GLSM acquire
large vevs.     

A similar analysis can be carried out for this model with all 
three tachyons in eq. (\ref{threetachyons}) turned on. This is
straightforward and we will not reproduce the results here. 
Let us however point out that there is an important difference
in calculating the phases of the GLSM with more than two $U(1)$
gauge fields. Namely, there might be extra columns 
in the charge matrix corresponding to the intersection of convex 
hulls obtained using triplets of the original field vectors \cite{mn}.  
These can be obtained from our analysis by studying the Lagrangian
(for the three parameter case) with two of the six fields acquiring
large vevs at a time, and then computing the D-term equations for
the reduced Lagrangian (obtained by suitable linear combinations of
the D-term equations of the original GLSM). In this particular 
example, one can check that out of the possible $15$ 
combinations where two of the six original fields in
eq. (\ref{threetachyons}) take large vevs, eight yield independent
Lagrangians, and writing down the D-term equations for these, 
we get precisely the points listed in \cite{mn}. We have checked
that this holds for generic examples. The entire information of
the phases are contained in the GLSM Lagrangian, and can be obtained
using our methods.  

Finally, we will briefly focus on Abelian 
non-cyclic orbifolds of $\BC^3$, of the form $\BC^3/\BZ_m\times \BZ_n$.
The analysis of the previous section can be applied here, once
we write down the charge matrix of the GLSM appropriate to the singularity.   
We will start here with the simplest example of such a singularity that
has tachyons in the spectrum : the orbifold $\BC^3/\BZ_2\times \BZ_5$
where the orbifolding action is given by
\begin{eqnarray}
&~& g_1~:~\left(Z^1,Z^2,Z^3\right) \to 
\left(-Z^1,-Z^2,Z^3\right)\nonumber\\
&~& g_2~:~\left(Z^1,Z^2,Z^3\right) \to
\left(Z^1,\omega Z^2, \omega^2 Z^3\right)
\end{eqnarray}
where $\omega = e^{\frac{2\pi i}{5}}$. There are four twisted sectors
here, with fractional R-charges $\left(\frac{1}{2},\frac{1}{2},0\right)$,
$\left(0,\frac{1}{5},\frac{2}{5}\right)$, 
$\left(0,\frac{3}{5},\frac{1}{5}\right)$, 
$\left(\frac{1}{2},\frac{1}{10},\frac{1}{5}\right)$, of which 
the first is marginal and the others are relevant. We will illustrate
our method of the previous section by focussing on a single tachyon,
i.e we consider turning on the twisted sectors corresponding to the
R-charges $\left(\frac{1}{2},\frac{1}{2},0\right)$ and
$\left(0,\frac{1}{5},\frac{2}{5}\right)$. In this case, the GLSM charge
matrix calculated from the toric data is
\begin{equation}
Q = \pmatrix{1&1&0&-2&0\cr 0&1&2&0&-5}
\end{equation}
where the first D-term is supersymmetric. Using the Lagrangian in 
eq. (\ref{lagtwopargen}), we can compute the sigma model 
metrics (and the phases) of this theory. For example, the unresolved
orbifold has the metric \footnote{We use the same notation as the 
previous section}
\begin{equation}
ds_{45}^2 = \sum_{i=1}^3 d\rho_i^2 + \frac{\rho_1^2}{(2)^2}d\tilde{\theta_1}^2
+\frac{\rho_1^2}{(10)^2}d\tilde{\theta_2}^2+
\frac{\rho_3^2}{(2)^2}d\tilde{\theta_3}^2  
\end{equation}
where the $\rho_i$ are the real parts of $\phi_i$ and the gauge invariant
angles are now $\tilde{\theta_1}=2\theta_1+\theta_4$,
$\tilde{\theta_2}=10\theta_2+5\theta_4+2\theta_5$,
$\tilde{\theta_3}=5\theta_3+2\theta_5$. Most of the phases of this
model are flat. In the region of moduli space where $\phi_1$ and
$\phi_5$ acquire large vevs (this is a distinct phase of the theory),
we obtain the metric
\begin{equation}
ds_{15}^2=d\rho_2^2 + d\rho_3^2 + d\rho_4^2 + 
\frac{\rho_2^2}{(5)^2}d\tilde{\theta_2}^2 +
\frac{\rho_3^2}{(5)^2}d\tilde{\theta_3}^2+
\rho_4^2d\tilde{\theta_4}^2
\end{equation}
where now $\tilde{\theta_2}=5(\theta_1-\theta_2)+\theta_5$,
$\tilde{\theta_3}=5\theta_3+2\theta_5$,
$\tilde{\theta_4}=2\theta_1+2\theta_4$,
which is recognised to be the metric for the space
$\BC^2/\BZ_{5(2)} \times \BC$. Similarly in the region of moduli
space where $\phi_1$ and $\phi_3$ acquire large vevs, the 
metric is seen to be that of $\BC^2/\BZ_{2(1)} \times \BC$. 
This is the same metric seen in the region of moduli space
where $\phi_3$ and $\phi_4$ acquire large vevs. The latter is
of course a distinct phase of the theory as can be seen
by plotting the charge vectors. 

We conclude by briefly commenting on the orbifold 
$\BC^3/\BZ_5\times \BZ_5$. We consider the two-tachyon system
with the relevant twisted sectors with R-charges 
$\left(\frac{1}{5},\frac{2}{5},0\right)$ and 
$\left(0,\frac{1}{5},\frac{2}{5},0\right)$ being turned on. 
The GLSM charge matrix in this case can be seen to be
\begin{equation}
Q = \pmatrix{1&2&0&-5&0\cr0&1&2&0&-5}
\end{equation}
It is straightforward to compute the five distinct phases in this
example. In the unresolved phase, the classical sigma model metric
is given by 
\begin{equation}
ds_{45}^2=\sum_{i=1}^3d\rho_i^2 + 
\frac{\rho_1^2}{(5)^2}d\tilde{\theta_1}^2+
\frac{\rho_2^2}{(5)^2}d\tilde{\theta_2}^2+
\frac{\rho_3^2}{(5)^2}d\tilde{\theta_3}^2+
\end{equation}
Where $\tilde{\theta_1}=5\theta_1+\theta_4$,
$\tilde{\theta_1}=5\theta_2+2\theta_4-\theta_5$,
$\tilde{\theta_3}=5\theta_3+2\theta_5$. When, for example, the fields
$\phi_1$ and $\phi_5$ take very large vevs, it is easy to see that
the metric for this region is that of $\BC^2/\BZ_{5(2)} \times \BC$.

Of course in order to understand the possibilities of flop transitions
etc. in these generic orbifolds, one needs to further analyse this
system including all the tachyons. We will leave such an analysis
to a future publication. 

\section{Conclusions}

In this paper, we have carried out a GLSM analysis of generic orbifolds
of $\BC^2$ and $\BC^3$. We have constructed the most general form of 
the GLSM Lagrangian, entirely in terms of the toric data for the
singularity which the model describes, from which the phases for these 
models can be read off. Our methods, which are applicable to GLSMs with
arbitrary charge matrices generalise the results of \cite{mn},
\cite{narayan}. These can, in particular, be used to study the phases
of generic GLSMs in lines with \cite{narayan} which may not have 
an SCFT description.    

As we have pointed out in section 3, it would be interesting to 
use the expression for the general multi parameter GLSM that we have 
derived in (\ref{lagrpargen}), to study D-brane dynamics in 
these models, in the lines of \cite{hiqv},\cite{tapo3}. In these 
papers, single parameter GLSMs were used to study supersymmetric
Calabi-Yau target spaces. In particular, in \cite{tapo3}, D-branes
in the non-linear sigma model limit of the open string GLSMs 
(i.e GLSMs with boundaries) were studied 
for the single parameter case, and was already seen to give rise 
to some interesting physics. The same analysis can be carried out 
using our results for general GLSMs with boundaries, 
with arbitrary number of gauge
groups and arbitrary charges. The behaviour of D-branes in various 
phases of these models can be studied using the results here, and is
expected to yield a rich structure. Further, we can also study 
D-branes in non-supersymmetric backgrounds using these. In particular,
the roles of fermions in these theories (with consistent Type II
GSO projection) need to be understood. We leave such a detailed study
for the future. 

Finally, it would be interesting to understand the full phase 
structure of generic orbifolds of the form $\BC^3/\BZ_m\times\BZ_n$, 
and possibilities of flop transitions therein, from the 
analysis that we have presented here. This can be carried out in
a straightforward manner from the analysis presented here.  

\vspace{1cm} 
\noindent
{\bf Acknowledgments}\\

\noindent
We would like to thank S. Govindarajan for private communications. Thanks
are due to S. Mukhopadhyay and K. Ray for helpful email correspondence, 
and to K. Narayan for an interesting discussion.

\end{document}